\documentstyle[psfig,nicV]{article}
\def\c12ag{$^{12}C(\alpha,\gamma)^{16}O$}
\def\n16{$^{16}N$}
\begin{document}
%
%Begin Heading
%
\heading{%
How Well Do We Know the Beta-Decay of $^{16}$N and Oxygen 
Formation in Helium Burning?\\
%
%End Heading
}
\par\medskip\noindent
%
%Begin Author names
\author{%
Ralph H France III$^{1,2}$, Moshe Gai$^{1,2}$
%End Author names
}
%First address
\address{
Dept of Physics U-46, University of Connecticut, 2152 Hillside Rd. 
Storrs, Ct. 06269-3046, USA.
}
% Second Address
\address{
Wright Nuclear Structure Laboratory, PO Box 208124 ,Yale University, 
272 Whitney Ave, New Haven, Ct 06520-8124, USA.
}
%\address{%
% Third Address
% Here
%
%}
\begin{abstract}
Contrary to claims that the problem has been solved, 
the astrophysical E1 S-factor of \c12ag
is not yet well known.  R-Matrix analyses of elastic 
scattering data,\c12ag data, and
data on the beta-delayed alpha-particle emission of 
$^{16}N$ are not consistent and a small 
S-factor solution [$S_{E1}(300)$] cannot be ruled out.  
In particular, data on the beta-delayed alpha-particle 
emission of $^{16}N$ do not agree.  The unaltered Mainz 
data do not agree with TRIUMF, but agree with 
both Seattle and Yale-UConn.  The TRIUMF collaboration 
has recalibrated
the Mainz('71) data; however,
we dispute both the alteration of the Mainz data performed 
by the TRIUMF collaboration and the very justification 
for the recalibration.
\end{abstract}
\section{Introduction}
Recently, a measurement of the beta-delayed alpha-particle 
emission of $^{16}N$ was performed at TRIUMF
\cite{Az94,Az97,Bu93}, which, together with an R-Matrix 
analysis of these and related data, was used to 
extract a value for the p-wave astrophysical S-factor 
of the \c12ag reaction.
Such an analysis relies upon
accurate knowledge of the line-shape of the spectrum of 
the beta-delayed
alpha-particle emission of \n16.
In the same paper \cite{Az94} a comparison
with the Mainz('71) data is shown, as communicated
to Dr. F.C. Barker by Dr. H. W\"{a}ffler \cite{Ba96} 
and published \cite{Ha69,Ha70,Ne74}, 
and it is claimed \cite{Az94} that the Mainz('71) 
spectrum "...is difficult to fit..."
due to a broader line-shape.  Hence the Mainz('71) data have been
largely ignored by these and other authors.  In this papert we
demonstrate the validity of the original Mainz('71) calibration,
 with which two additional experiments
\cite{Fr97,Zh95} agree quite well, see Fig. 1.

\section{The Recalibration of the Mainz('71) Spectrum}
\begin{figure}
\centerline
{\vbox{\psfig{figure=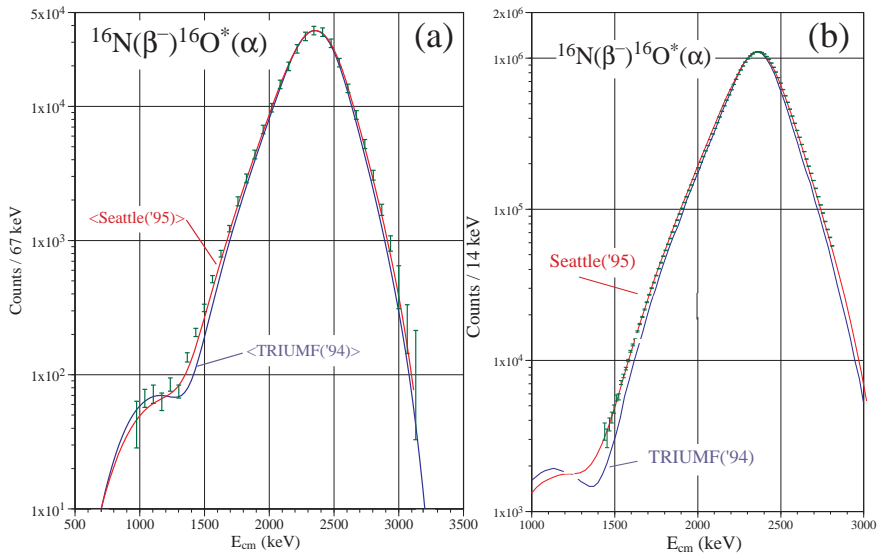,height=8.2cm}}}
\caption[]{(a) The TRIUMF('94) and Seattle('95) {\bf R}-Matrix 
fitted curves compared 
with the Yale-UConn('96) data, corrected for line shape 
\cite{Fr97}. (b) TRIUMF('94) and 
Seattle('95) {\bf R}-Matrix fitted curves compared
with the unaltered Mainz('71) data set.
}
\end{figure}

While the unaltered Mainz('71) data disagree 
with the TRIUMF('94) experiment on both the high and low
energy sides of the broad $1^-$ peak, the TRIUMF 
group \cite{Az94,Az97} has produced a recalibration of 
the Mainz spectrum leading to a very different 
spectrum with "The difference ranges from 6.5 keV at 
the low end to 18 keV at the higher energies" \cite{Po96}. 
The altered Mainz data agree with the later 
TRIUMF data \cite{Az94} on the high energy side, but
disagree even more significantly on the low energy
 side.  It is claimed "...the Mainz spectrum shows evidence of an 
enhancement on the low energy side of the peak that is 
likely to be the result of the low energy tail of the 
system response function.  Hence the Mainz('71) data
have been considered faulty.

Azuma {\em et al.} \cite{Az97} make the 
claim that the energy calibration (10.60 kev/ch) 
contained in W\"{a}ffler's communication 
to Barker \cite{Ba96} is wrong, and that the Mainz('71) 
spectrum can be self-calibrated with high accuracy.  
They use W\"{a}ffler's statement (in his letter) that
"...channel 37 corresponds to 1281 keV..." 
\cite{Ba96} and claim that the centroid 
of the $2^+$ state is accurately extracted
from the Mainz('71) spectrum \cite{Ba96}.  Using only the 
well known energy of the $2^+$  
state they derive a different energy dispersion (
10.45 keV/ch \cite{Ba96}).

At first we note that it seems arbitrary that Azuma {\em et al.}
\cite{Az97} adopt part of W\"{a}ffler's calibration (channel 37
is 1281 keV) \cite{Ba96}, but reject the very dispersion 
(10.60 keV/ch) used to calibrate it.
Even accepting this, their recalibration \cite{Az97} 
depends upon the ability to precisely extract the centroid
of the $2^+$ state in the Mainz('71) spectrum \cite{Ba96}.
In Fig. 2a we show the Mainz('71) data over the region of interest. 
The raw data show a very strong 
energy dependence, and in the vicinity of channel 106 one observes
a minuscule excess of counts, most likely due to a 
contribution from the $2^+$ state. 
An accurate extraction of a
centroid for this excess is very dependent on the choice of background
and requires data with extremely good statistics. The exact energy dependence
of the background cannot be calculated {\em ab initio} as it is a convolution
of the beta-decay phase space with contributions from the broad $1^-$
state plus non-calculable background states.

\begin{figure}
\centerline
{\vbox{\psfig{figure=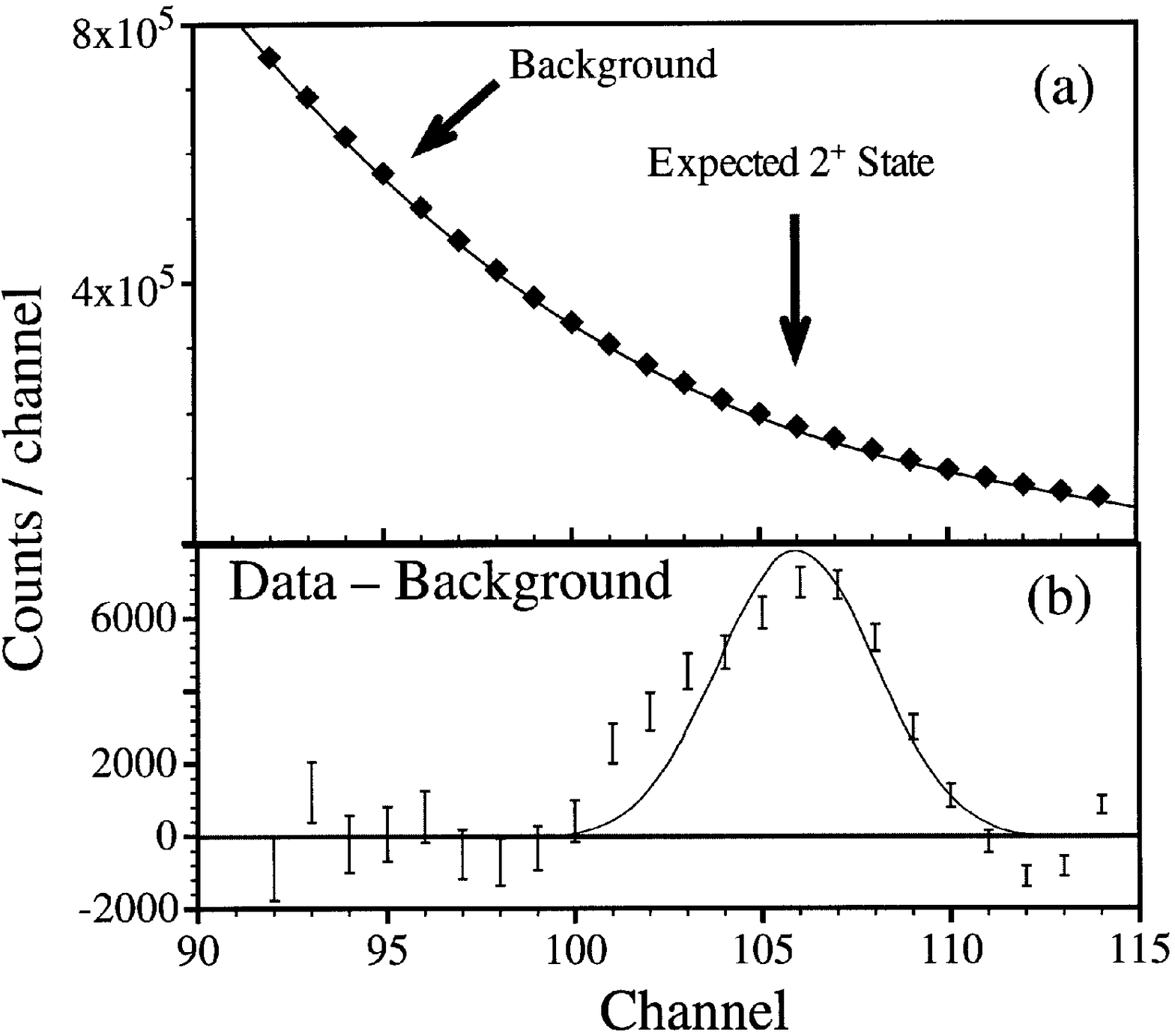,height=6.5cm}}}
\caption[]{(a) The Mainz('71) data with a background fit, (b) and 
subtracted from the data with a gaussian fit.  This gaussian 
fit for the expected $2^+$ state is consistent
with the original Mainz('71) calibration.
}
\end{figure}

\begin{figure}
\centerline
{\vbox{\psfig{figure=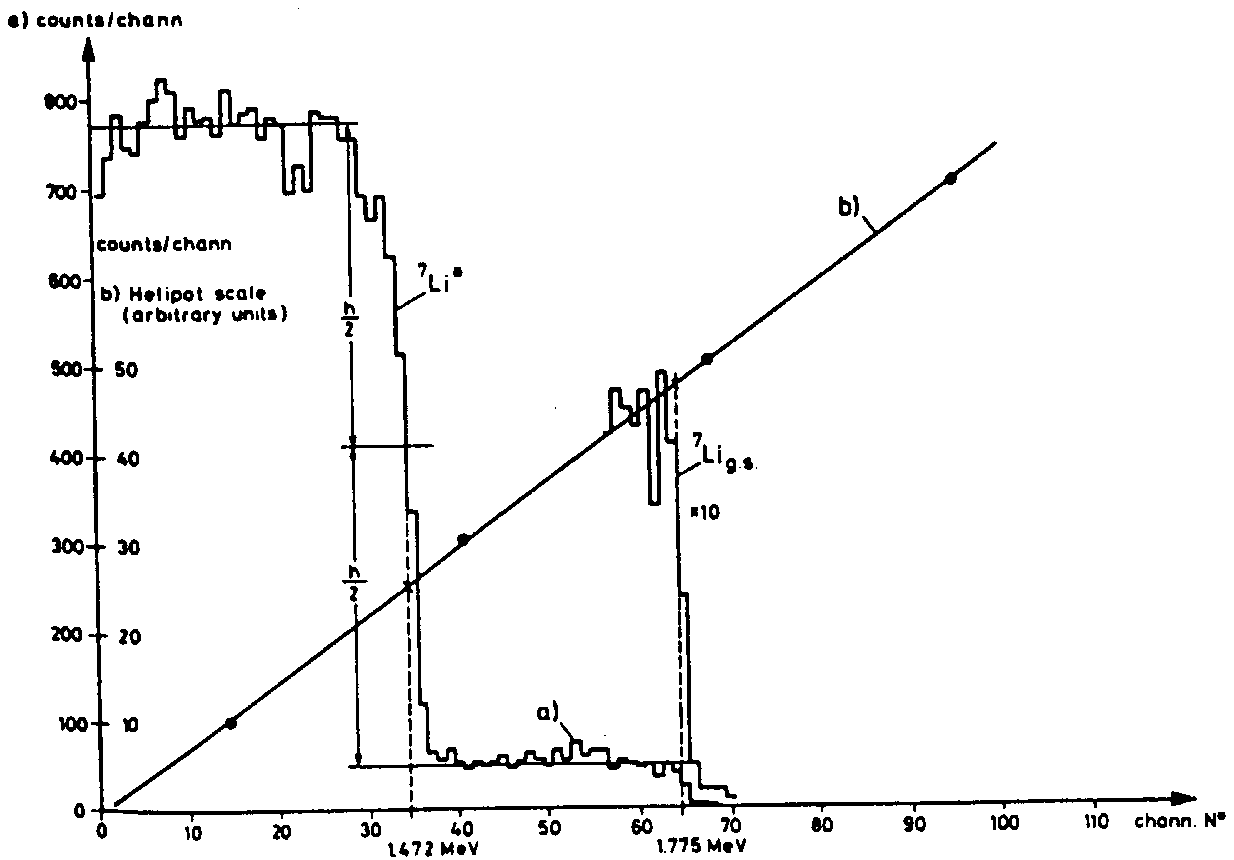,height=7.5cm}}}
\caption[]{An example of the original Mainz('71) 
calibration \cite{Ha69,Ha70,Ne74} with a signal 
to noise ratio of 12:1.  This should be compared to
the spectrum shown in Fig. 2 with a signal to background ratio of a few percent 
used by the TRIUMF group \cite{Az94,Az97} to recalibrate the Mainz data. 
}
\end{figure}

In Fig. 2a we show the original Mainz('71) data
 with a fourth order 
polynomial background fit ($\chi ^2 /\nu \ = \ 0.7$ 
for ch 92-100 and $\chi ^2 
/\nu \ = \ 5.2$ for ch 110-114).  The background 
subtracted data are shown in 
Fig. 2b together with a fit to a gaussian with a 
centroid fixed at the 
expected energy of the $2^+$ state 
($E_\alpha$ = 2.0115 MeV expected 
at channel 105.9).  The resultant fit is not 
inconsistent with 
the expected shape of the 
$2^+$ state when allowing for undulations in the background.  
This fit casts a strong doubt on
the necessity of the recalibration.  In addition, we emphasize 
that the choice of background leads to a systmatic uncertainty in the 
extracted centroid (ch 105.2 to 107.5)
greater than the correction introduced in Ref. \cite{Az94,Az97,Bu93}. 
 We emphasize that this 
is a systematic uncertainty and thus different 
from statistical uncertainties
(derived from chi-square considerations).

In Fig. 3 we show the original calibration data from 
the Mainz('71) experiment \cite{Ha69,Ha70,Ne74}.  
using the $^{10}B(n,\alpha)^7Li$ procedure which 
resulted in an energy uncertainty of $\pm 10 keV$.  In
contrast to the spectrum used for the recalibration with a signal a few
percent above background, the 
original calibration spectrum has a signal to noise 
ratio of 12.  We strongly doubt the stated accuracy of the recallibration
procedure as given by the TRIUMF collaboration.  While they use data with 
a signal to background ratio nearly 1000 times worse (i.e. 12:1 vs about 1\% 
above background see Figs. 2 and 3)  the TRIUMF collaboration claims to 
extract a centroid with a factor of 5 better precission ($\pm2$ keV vs.
 $\pm10$ keV).

\section{Conclusion}
  
It is most important to evaluate the effect of the various data sets on
the extracted p-wave astrophysical S-factor of the \c12ag reaction.
This question is beyond the scope of this short
contribution, but we remark that the \n16 spectrum allows for extracting
the reduced alpha-particle width of the bound $1^-$ state at 7.12 MeV,
but it can not determine a priori 
whether the interference between the bound and quasi-bound $1^-$ states
is constructive or destructive, and cannot
rule out the small S-factor solution (i.e $S_{E1}<20$ keV-b) \cite{Ha97}. Clearly 
a change in the line shape by as much as a factor of two at 1.4 MeV 
(the region of the interference minimum, see Fig. 1 
is expected to, for example, significantly
alter the f-wave contribution and thus the
extracted p-wave astrophysical S-factor of the \c12ag reaction.

\acknowledgements{We wish to acknowledge communications
from Fred C. Barker and James Powell, and discussions
with James D. King.}

\vfill
\end{document}